%% file: draft_v1.tex
\documentclass[12pt]{article}
\usepackage[a4paper, total={7in, 9in}]{geometry}
\input{setup}

\usepackage{authblk}

\title{Boundary quenches in (1+1)-dimensional \\ conformal field theory}
\author[1,2]{Michele Fossati}
\author[3]{Colin Rylands}
\author[4]{Eytan Grosfeld}
\author[5]{Eran Sela}
\author[1]{Pasquale Calabrese}

\date{\today}

\affil[1]{\textit{SISSA and INFN, Via Bonomea 265, 34136 Trieste, Italy}}
\affil[2]{\textit{Southern Denmark University, Campusvej 55, 5230 Odense, Denmark}}
\affil[3]{\textit{Centre for Fluid and Complex Systems, Coventry University, Coventry, CV1 2TT, United Kingdom}}
\affil[4]{\textit{Department of Physics, Ben-Gurion University of the Negev, Be’er-Sheva 84105, Israel}}
\affil[5]{\textit{School of Physics and Astronomy, Tel Aviv University, Tel Aviv 6997801, Israel}}

\begin{document}
\maketitle
\abstract{
We investigate a class of local quantum quenches in which the conformal boundary condition of a $(1+1)$-dimensional conformal field theory is abruptly changed. We derive a remarkably simple and universal expression for the time evolution of one-point functions on the half-line. This result provides a direct description of the propagation of the disturbance generated by the quench and, in turn, allows us to determine the dynamics of bipartite entanglement for subsystems adjacent to the boundary. We show that, once the subsystem becomes fully causally connected to the quench event, the entanglement entropy undergoes a sharp finite jump whose magnitude is universally given by the logarithm of the ratio of the boundary $g$-factors associated with the initial and final boundary conditions. We benchmark these analytical predictions against Matrix Product State simulations of the critical Ising spin chain, finding excellent agreement. The numerical analysis further allows us to investigate the time evolution of the spin-flip entanglement asymmetry, revealing how the symmetry-breaking perturbation emitted from the boundary propagates through the system. Our results uncover universal dynamical signatures of boundary quenches and establish a direct connection between nonequilibrium entanglement dynamics and boundary critical phenomena. 

}

\newpage

\tableofcontents

\section{Introduction}

A \emph{quantum quench} is a nonequilibrium protocol in which the ground state (or, more generally, an eigenstate) of a Hamiltonian is suddenly evolved under a different, non-commuting Hamiltonian~\cite{Calabrese:2006rx,Calabrese:2007rg}. Such protocols are of considerable experimental interest, particularly in systems of ultracold atoms and trapped ions, which can be sufficiently well isolated from the environment to be accurately described as closed quantum systems undergoing unitary time evolution. Moreover, the rapid tuning of Hamiltonian parameters required to implement quenches is routinely achieved in these platforms~\cite{Kinoshita:2006xby,Hofferberth:2007qfs,Trotzky:2011flm,Gring:2012usl,Cheneau:2012zdz}.
On the theoretical side, $(1+1)$-dimensional conformal field theories (CFTs) provide a particularly powerful framework for studying quantum quenches, owing to the wealth of analytical techniques that often allow for exact solutions. This approach has proved remarkably fruitful, leading to universal insights into nonequilibrium quantum dynamics that extend well beyond conformal systems. A notable example is the quasiparticle picture for the spreading of entanglement following a quantum quench, which was originally formulated within CFT~\cite{Calabrese:2005in} and later generalized to interacting integrable models~\cite{Alba:2017ekd}.

A fundamental distinction is between \emph{global} and \emph{local} quantum quenches. For a local quantum system, whose Hamiltonian is the spatial integral of a local Hamiltonian density (or, in lattice models, a sum of local terms), a \emph{global quench} consists of a sudden change of the Hamiltonian density throughout the entire system. By contrast, in a \emph{local quench} the modification is confined to a finite spatial region or even a single point~\cite{Calabrese:2007mtj,Eisler_2007,Eisler:2008jau,Kleine_2008,Igloi:2009xku,Hsu:2009bdm,Polkovnikov:2010yn,Cardy:2011zz,Eisler_2012,Collura:2013kag,Asplund:2013zba,Wen:2015qwa,Dahan_2017,Feldman:2019upn, Dahan:2019duw,DiGiulio:2021noo,Kudler-Flam:2020xqu,Capizzi:2022sfk,Rylands:2023duu,Kudler-Flam:2023ahk,Caputa:2025dep,bonsignori2025},  generating a localized excitation whose subsequent propagation governs the nonequilibrium dynamics.

Among the most studied local quenches is the  \emph{cut and glue} protocol~\cite{Calabrese:2007mtj, Calabrese:2008gsv, Stephan:2011kcw, Calabrese:2016xau, Bertini:2016tmj}. In this case, the initial state is the ground state of a certain homogeneous Hamiltonian on two disjoint regions, $(-\infty, -\epsilon]$ and $[\epsilon, \infty)$, with $\epsilon$ a small positive number, i.e., the real line is initially cut in half. Then the two boundaries at $-\epsilon$ and $\epsilon$ are glued together, forming a uniform Hamiltonian on the real line, and the initial state is evolved in time. 
The quench event can be thought of as the removal of a factorizing defect (like the ones studied in~\cite{Popov:2025cha}).
In another well studied class of quench protocols, the initial state is prepared by applying a genuine point operator to the ground state, thus creating a low energy excited state.
The state is then evolved in time with the ordinary uniform Hamiltonian~\cite{He:2014mwa, Zhang:2019kwu,Guo:2022sfl, He:2023eap, Krasznai:2024qnk}.  
Despite both of them being local quenches, the two protocols display a different phenomenology. For instance, in CFTs, the entanglement entropy over a bipartition has a logarithmic growth in time for the cut and glue protocol, while in the operator quench it  undergoes a sharp jump in time (equal, in the case of a rational CFT, to the logarithm of the quantum dimension associated to the operator inserted~\cite{He:2014mwa}).

In this paper, we investigate a different class of local quantum quenches, namely \emph{boundary quenches}. The system is initially prepared in the ground state of a $(1+1)$-dimensional theory on a spatial manifold with a boundary (either the half-line or a finite interval), with prescribed boundary conditions. At time $t=0$, one of the boundary conditions is suddenly changed (quenched) while the bulk Hamiltonian remains unchanged. The subsequent unitary evolution is therefore entirely driven by the spreading of a localized perturbation at the boundary, making boundary quenches a particularly simple and universal realization of a local quench. 

We study analytically one-point functions in the case in which the spatial manifold is a semi-infinite interval (i.e. when there is only one boundary condition), finding that they follow a light-cone behavior: if the point is space-like separated from the quench event (the change in the boundary condition), one-point functions are the same as in the ground state with the initial boundary condition; if instead the point is time-like separated from the quench event, they are the same as in the ground state with the quenched boundary condition.
The study of one-point functions allows us also to derive the time evolution of the entanglement entropy for a bipartition in which one of the subsystems is adjacent to the boundary. We find a jump equal to $\log(g_b / g_a)$ ($g_a$ and $g_b$ being respectively the $g$-factors \cite{Affleck:1991tk} of the initial and final boundary condition) in the entanglement entropy, happening at the time when the endpoint of the subsystem becomes causally connected to the quench event. 

From these results, it is apparent that the boundary quench combines features of both of the previously discussed protocols. On the one hand, as in the \emph{cut and glue} protocol, the quench is realized by modifying an extended object in the theory, since the change in the boundary condition is analogous to removing a factorizing defect. On the other hand, its dynamical phenomenology is much closer to that of a local operator insertion. Indeed, the entanglement entropy of a fixed bipartition undergoes a sharp finite jump at a specific time.

After the general analysis,
we specialize to the Ising CFT and benchmark our analytical predictions against numerical simulations of the critical Ising spin chain, whose low-energy physics is described by the Ising CFT. The simulations are performed on finite chains using Matrix Product State (MPS) techniques. We find excellent agreement with the predicted behavior of both the one-point functions and the entanglement entropy on length scales smaller than the system size, where the infinite-chain approximation is expected to hold.
The observed dynamics can be interpreted in terms of a localized excitation propagating at the speed of sound and scattering off the system boundaries. Furthermore, we provide a simple algebraic argument that predicts whether the magnetization changes sign at the time corresponding to twice the interval length (in units of the sound velocity), thereby explaining the different behaviors observed for the various boundary conditions.

Additionally, we numerically compute the second Rényi \emph{entanglement asymmetry} associated with the $\mathbb{Z}_2$ symmetry of the spin chain. Entanglement asymmetry is a quantum-information measure of symmetry breaking~\cite{Ares:2022koq,Casini:2019kex,Casini:2020rgj,Vaccaro:2007drw,Gour:2009hng}. It is defined by comparing the reduced density matrix of a subsystem with its symmetrized counterpart, thereby quantifying the degree of symmetry breaking at the level of the subsystem.
It has been shown to be an ideal proxy for the quantum Mpemba effect \cite{Ares:2025onj,Calabrese_rev_2026} and it is experimentally measurable \cite{Joshi:2024sup,Yang2026}.
We evaluate the second Rényi entanglement asymmetry for subsystems adjacent to either the left or the right boundary as a function of time, and use it to characterize the propagation of symmetry breaking induced by the boundary quench.

The paper is organized as follows. In~\cref{sec:time-evol-1pt-funcs} we present the analytical results regarding one-point functions and entanglement entropy in the semi-infinite geometry in a general CFT. In~\cref{sec:ising} we study in detail the example of the Ising CFT, with numerical simulations performed on the critical Ising chain.

While this manuscript was being completed, Ref.~\cite{Bajnok:2026dag} appeared on the arXiv. It studies boundary quenches in conformal field theories and their integrable deformations, with a particular focus on the Yang--Lee CFT. The main results of the present work, namely~\cref{eq:one-pt-func-result,eq:entropy-jump}, the analysis of the Ising CFT, and the numerical simulations of the critical Ising chain, are, to the best of our knowledge, original and do not overlap with those of Ref.~\cite{Bajnok:2026dag}.

\section{Time evolution of one-point functions}
\label{sec:time-evol-1pt-funcs}
We study the time evolution of a $(1+1)$-dimensional CFT on the positive half-line $[0, +\infty)$. We consider the ground state with a certain boundary condition $a$ at $x = 0$ and evolve it with a different boundary condition $b$, i.e. we perform a local quench on the boundary condition. We assume that the boundary conditions is \emph{simple} in the sense of \cite{Choi:2023xjw}.

\begin{figure}
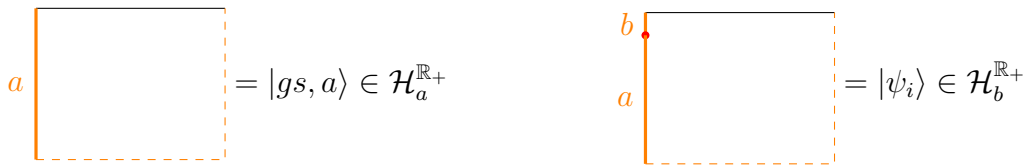

	\centering
	\begin{align*}
		\tikzmath[baseline=-1em]{
			\draw[] (0,0) -- (2.5,0);
			\draw[very thick, orange] (0,0) -- node[midway, left] {$a$} (0,-2);
			\draw[dashed, orange] (0,-2) -- (2.5,-2) -- (2.5,0);
		}
		&=  \ket{gs, a} \in \cH_a^{\bR_+}
		&
		\tikzmath[baseline=-1em]{
			\draw[] (0,0) -- (2.5,0);
			\draw[very thick, orange] (0,0) -- node[midway, left] {$b$} (0,-0.3);
			\fill[red] (0,-0.3) circle (1.5pt);
			\draw[very thick, orange] (0,-0.3) -- node[midway, left] {$a$} (0,-2);
			\draw[dashed, orange] (0,-2) -- (2.5,-2) -- (2.5,0);
		}
		&= \ket{\psi_i} \in \cH_b^{\bR_+}
	\end{align*}
	\caption{Euclidean path-integral preparation of the ground state with \bc $a$ (on the left), and path integral preparation of the initial state $\ket{\psi_i}$ of \eqref{eq:initial-state-quench} (on the right).}
	\label{fig:initial-state-quench}
\end{figure}

The ground state with boundary condition $a$ is denoted as $\ket{\gs, a}$ and is prepared by a euclidean path integral from time $\tau=-\infty$ to  $\tau=0$, and spatially extended on $[0, +\infty)$, see the left panel of \cref{fig:initial-state-quench}. It belongs to the Hilbert space $\cH_a^{\bR_+}$ of states on the positive half-line with \bc $a$. 

In order to evolve the state  $\ket {\gs, a} \in \cH_a^{\bR_+}$ with the time evolution in which the \bc $b$ is set, one has to first to embed $\ket{\gs, a}$ into $\cH_b^{\bR_+}$. This can be done with the use of a boundary-changing operator $\hat \phi: \cH_a^{\bR_+} \to \cH_b^{\bR_+}$, which we assume to be a primary.
The \emph{initial state}, the state at time $t=0$, is then 
\begin{equation}\label{eq:initial-state-quench}
	\ket{\psi_i} =
	e^{-\epsilon H_b} \hat \phi \ket{gs,a}  \quad \in \cH_b^{\bR^+}
\end{equation}
where $H_b$ is the Hamiltonian with boundary condition $b$ and $\epsilon >0$. The term $e^{-\epsilon H_b}$ with $\epsilon > 0$ is added such that the state has finite norm, as in \cite{Calabrese:2006rx} for global quenches, although we do not assume the state to have unit norm.
The dependence on $\epsilon$ will be removed by taking the limit $\epsilon \to 0$ at the end of the calculation. A pictorial description of this initial state preparation is in the right panel of \cref{fig:initial-state-quench}.

\begin{figure}
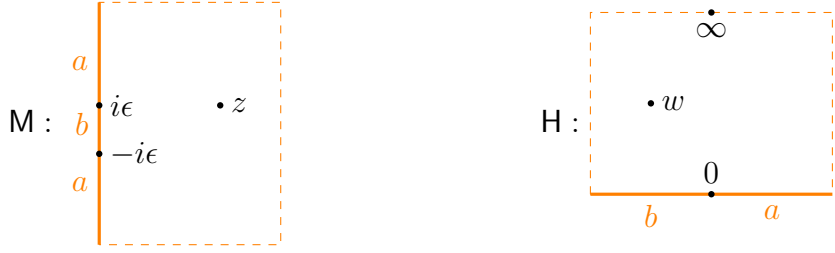

		\begin{align*}
				\sM &: 
				\tikzmath[scale=0.8]{
					\draw[dashed, orange] (0,-2) rectangle (3,2);
					\draw[very thick, orange] (0,-2) -- (0,2);
%
					\node[orange,left] at (0,0) {$b$};
					\node[orange,left] at (0,1) {$a$};
					\node[orange, left] at (0,-1) {$a$};
					\fill[] (0,0.3) circle (1.5pt) node[right]{$i\epsilon$};  
					\fill[] (0,-0.5) circle (1.5pt) node[right]{$-i\epsilon$}; 
					\fill (2,0.3) circle (1.5pt) node[right] {$z$};
				}
			&
			\sH &:
			\tikzmath[scale=0.8]{
				\draw[dashed, orange] (-2,-1) rectangle (2,2);
				\draw[very thick, orange] (-2,-1) -- (2,-1);
%
%
				\node[below,orange] at (-1,-1) {$b$};
				\node[below,orange] at (1,-1) {$a$};
				\fill[] (0,-1) circle (1.5pt) node[above] {$0$};
				\fill[] (0,2) circle (1.5pt) node[below] {$\infty$};
				\fill (-1,0.5) circle (1.5pt) node[right] {$w$};
			}
		\end{align*}
	\caption{Left: The geometry $\sM$, is the right half of the complex plane, with fields that change boundary condition from $a$ to $b$ and vice-versa inserted at $\pm i \epsilon$. Points on $\sM$ are denoted by $z = x + i \tau$. Right: the geometry $\sH$ is the upper half-plane, with boundary condition $a$ on the positive real axis, and boundary condition $b$ on the negative real axis. Boundary changing fields are at $0$ and at $\infty$. Points on $\sH$ are denoted by $w$.}
	\label{fig:one-point-function}
\end{figure}

We  consider the expectation value of a local spinless (i.e. with equal holomorphic and antiholomorphic conformal dimensions, $ h_\mathcal{O}=\bar h_\mathcal{O}$) primary operator $\hat{\mathcal O}(x) \in \End(\cH_b)$, with $x > 0$, on the state $\ket \psi$ at time $t > 0$
\begin{equation}
		\langle  \hat{\mathcal O}(x) \rangle_t = \frac{\bra{\psi_i} e^{itH_b} \hat{\mathcal O}(x) e^{-itH_b} \ket{\psi_i}}{\braket{\psi_i|\psi_i}} \, .
\end{equation}
We analytically continue time to Euclidean signature ($\tau = it$), obtaining 
\begin{equation}
	\langle \hat \cO(x) \rangle_\tau =  \frac{ \bra{\gs, a} \hat \phi^\dag e^{-(\epsilon - \tau) H_b} \hat \cO(x) e^{-(\epsilon+\tau) H_b} \hat \phi \ket{\gs, a} }{\bra{\gs, a} \hat{\phi}^\dag e^{-2\epsilon H_b} \hat{\phi} \ket{\gs, a}} \,.
\end{equation}
This can be seen as a ratio of a one-point function with a partition function. For both, the geometry is the right half of the complex plane, with boundary condition $a$ on the imaginary axis on $(-i \infty, -i \epsilon) \cup (i \epsilon, i \infty)$, and boundary condition $b$ on $(-i \epsilon, i \epsilon)$; boundary changing fields are inserted at $\pm i \epsilon$. We call this geometry $\sM$, represented in the left of \cref{fig:one-point-function}. In the numerator, we have the same geometry, with the insertion of the field $\cO$ at $z = x + i \tau$. The ratio is then equal to the one-point function of $\cO(z)$ on the geometry $\sM$
\begin{equation}
	\langle \hat \cO(x) \rangle_\tau = \langle \cO(z) \rangle_\sM \, .
\end{equation}



\noindent The geometry  $\sM$ can be mapped to the upper half-plane, $\sH$, by the global conformal map,
\begin{equation}
	\begin{aligned}
		f &: \sM \to \mathsf \sH  ~,\\
		f(z) & = \frac{z + i \epsilon}{z - i \epsilon} =: w~,
	\end{aligned}
\end{equation}
where, on $\sH$, the boundary conditions $a$ and $b$ apply in the regions $(0, +\infty)$ and $(-\infty, 0)$ respectively, while the boundary changing operators are inserted at $w=0$ and $w=\infty$ (see~\cref{fig:one-point-function} right). As a result, the one-point function of the scalar primary can be expressed as
\begin{equation}
	\langle \cO(z) \rangle_\sM = \abs{ f'(z) }^{\Delta_\cO}  \langle \cO(f(z)) \rangle_\sH,
\end{equation}
where $\Delta_\mathcal{O}=h_\mathcal{O}+\bar{h}_\mathcal{O}$ is the scaling dimension of $\mathcal{O}$.

Now recall that the one-point function of a bulk spinless field on  $\sH$ has a generic form
\begin{equation}
	 \langle \cO(w) \rangle_\sH = \cF_{ab}(\cos \theta) ( 2\Im(w) )^{-\Delta_\cO},
\end{equation}
for some function $\mathcal{F}_{ab}$ whose explicit form is not needed and where $\theta = \arg w$ \cite{Burkhardt:1991eq}. 
When $w$ is close to the real axis, the one-point function feels only the presence of the boundary condition $a$ or $b$ respectively. Accordingly,  $\cF_{ab}(1) = \cA^{a}_{\cO, \bI}$ and $\cF_{ab}(-1) = \cA^{b}_{\cO, \bI}$, where $\cA^{a (b)}_{\cO, \varphi}$ is the bulk-boundary structure constant between the bulk field $\cO$ and a boundary field $\varphi$ on the boundary $a(b)$, and $\bI$ is the identity operator on the boundary. Bulk-boundary structure constants are known for a wide class of CFTs, see e.g.~\cite{Runkel:1998he, Runkel:1999dz}. 

For the conformal map $f$, we have 
\begin{align}
	\Im(w) &= \frac{2 x \epsilon}{x^2 + (\tau - \epsilon)^2} & 
	|f'(z)| &= \frac{2 \epsilon}{x^2 + (\tau - \epsilon)^2},
\end{align}
so that 
\begin{equation}
	\langle \cO(z) \rangle_\sM = \cF_{ab} (\cos \theta) (2 x)^{-\Delta_\cO}.
\end{equation}
Then we have
\begin{equation}
	\cos \theta = \frac{\Re(w)}{\abs w} = \frac{x^2 + \tau^2 -\epsilon^2}{\sqrt{(x^2 + (\epsilon - \tau)^2)(x^2 + (\epsilon + \tau)^2)}},
\end{equation}
which, after setting $\tau = i t$ and sending $\epsilon \to 0$, becomes
\begin{equation}
	\lim_{\epsilon \to 0} \cos \theta \big |_{\tau = i t} = \sign(x-t).
\end{equation}
Hence we find that, in the limit $\epsilon \to 0$, the analytic continuation of $\cF_{ab}(\cos \theta)$ is equal to $\cA^a_{\cO, \bI}$ for $x > t$, and equal to $\cA^b_{\cO, \bI}$ for $x < t$.
Now recall that the expectation value of $\hat \cO(x)$ on the ground state with boundary condition $a$ is 
\begin{equation} \label{eq:one-pt-func-gs}
	\bra{\gs, a} \hat \cO(x) \ket{\gs, a} = \cA^a_{\cO, \bI} (2x)^{-\Delta_\cO}.
\end{equation}
Therefore, we get that for $x > t$ the expectation value of $\hat \cO(x)$ at time $t$ is the same as the one in the ground state with boundary condition $a$, while for $x < t$ it is equal to the one in the ground state with boundary condition $b$. 
\begin{equation}\label{eq:one-pt-func-result}
	\langle \hat O (x) \rangle_t = 
	\begin{cases}
		\bra{\gs, a} \hat \cO(x) \ket{\gs, a} & \text{if } x > t \\
		 \bra{\gs, b} \hat \cO(x) \ket{\gs, b} & \text{if } x < t .
	\end{cases}
\end{equation}
We stress that the state evolved at time $t>0$ is \emph{not} the ground state of the Hamiltonian with boundary condition $b$.

\subsection{Time evolution of the entanglement entropy}
The same construction can be used to study the entanglement entropy at time $t$, on a spatial bipartition where the half-line $[0,\infty)$ is split in two disjoint intervals $A = [0, x - \delta]$ and $B = [x+\delta, \infty)$. $\delta$ is a UV cutoff that has to be inserted to decompose the Hilbert space in the tensor product of the Hilbert space supported on $A$ and the one on $B$. Morally, the bipartition can be thought of as between $[0,x]$ and its complement.

Let $\rho(t) = \ket{\psi(t)} \!\bra{\psi(t)}$ be the density matrix of the state on the half-line at time $t$. To perform the partial trace, one has to first embed the space on the factorized Hilbert space. As explained in~\cite{Ohmori:2014eia} this is done with a map 
\begin{equation}
	\iota: \cH_b^{\bR_+} \to \cH_{(b,d)}^{A} \otimes \cH_d^B,
\end{equation}
which is defined by setting \bc $d$ on the \emph{entangling surface} $[x-\delta, x+\delta]$. In this case, the Hilbert space $\cH_{(b,d)}^A$ associated to $A$ is a finite interval, and hence has two boundary conditions. The density matrix on the factorized space is $\rho_{AB}(t) = \iota \, \rho(t) \, \iota^\dag$, and the reduced density matrix on $A$ is 
\begin{equation}
	\rho_A(t) = \Tr_B[\rho_{AB}(t)]. 
\end{equation}
The entanglement entropy $S_A(t) = -\Tr[\rho_A(t) \log \rho_A(t)]$ can be accessed from the $n$-th R\'enyi entropy  $S^{(n)}_A(t) = (1-n)^{-1} \log \Tr \slr{\rho_A(t)^n}$ and taking the limit $n \to 1$ \cite{Calabrese:2004eu}.

The R\'enyi entropy at time $t$ is then computed from the identity
\begin{equation} \label{eq:tr-rho-n-time}
	\Tr \slr{ \rho_A(t)^n } = {}^{\otimes n} \! \bra{\psi(t)} \iota^\dag \, R_A \, \iota \ket{\psi(t)}^{\otimes n},
\end{equation}
where 
\begin{equation}
	R_A: (\cH^A \otimes \cH^B)^{\otimes n} \to (\cH^A \otimes \cH^B)^{\otimes n}
\end{equation}
is the linear map that permutes cyclically the copies of $\cH^A$ (shifting their index by one), and acts as the identity on all the factors of $\cH^B$.

As before, we access the real time evolution by going to Euclidean signature, evaluating a correlator, and analytically continuing the result back.
Let $\Tr \slr{\rho_A(i\tau)^n}$ be the analytic continuation in Euclidean time of $\Tr \slr{\rho_A(t)^n}$. This can be expressed as the following ratio of partition functions (see \cref{fig:tr-rhoA-n}). In the numerator one has the partition function of $n$ copies of the CFT on the right half-plane, with boundary condition on the imaginary axis set to be $a^{\otimes n}$ on $(-i \infty, -i \epsilon) \cup (i \epsilon, i\infty)$ and $b^{\otimes n}$ on $(-i\epsilon, i \epsilon)$; a disk of radius $\delta$ centered in $z = x + i \tau$ is removed, and the boundary condition $d^{\otimes n}$ is set; the line operator $R$ which implements the permutation of the replicas connects the entangling surface with the boundary at $x=0$. 
Since the boundary conditions $a^{\otimes n}$ and $b^{\otimes n}$ are invariant under permutations of the replicas, not only is the line $R$ topological, but also the junction between $R$ and the boundary. In the denominator, we have almost the same partition function as the numerator, except that the line operator $R$ is omitted. The denominator comes from the normalization of the reduced density matrix $\rho_A(t)$. 

\begin{figure}
\begin{equation*}
	\Tr \slr{ \rho_A(i\tau)^n } = Z
	\lr{
		\tikzmath[scale=0.8]{
			\draw[dashed, orange] (0,-2) rectangle (2,2);
			\draw[very thick, orange] (0,-2) -- (0,2);
			
			\node[orange,left] at (0,0) {$b^{\otimes n}$};
			\node[orange,left] at (0,1) {$a^{\otimes n}$};
			\node[orange, left] at (0,-1) {$a^{\otimes n}$};
			
			\fill (0,0.3) circle (1.5pt);
			\fill (0,-0.3) circle (1.5pt);
			\draw[orange, very thick] (1,0) circle (3pt) node[above right] {$d^{\otimes n}$};
			\draw[purple] (0,0) -- node[midway, above] {$R$} (0.9,0);
		}
	}
	\Bigg / Z
	\lr{
		\tikzmath[scale=0.8]{
			\draw[dashed, orange] (0,-2) rectangle (2,2);
			\draw[very thick, orange] (0,-2) -- (0,2);   		
			\node[orange,left] at (0,0) {$b^{\otimes n}$};
			\node[orange,left] at (0,1) {$a^{\otimes n}$};
			\node[orange,left] at (0,-1) {$a^{\otimes n}$};
			
			\fill (0,0.3) circle (1.5pt);
			\fill (0,-0.3) circle (1.5pt);
			\draw[orange, very thick] (1,0) circle (3pt) node[above right] {$d^{\otimes n}$};
		}
	}.
\end{equation*}
\caption{The analytic continuation to euclidean time of \eqref{eq:tr-rho-n-time}, expressed in terms of partition functions of the $n$-replicated CFT.}
\label{fig:tr-rhoA-n}
\end{figure}

In the limit $\delta \to 0$ each circle with boundary condition $d$ appearing in the denominator becomes equal to the identity field times the $g$-factor $g_d$. The $n$ circles displayed in the numerator, which are twisted by $R$, become the replica twist field $\cT_n$ times $\delta^{\Delta_{\cT_n}}g_d$ (where $\Delta_{\cT_n} = \frac{c}{12} \lr{n - 1/n}$ is the scaling dimension of $\cT_n$ \cite{Calabrese:2009qy}). This is analogous to the procedure explained in  \cite{Ohmori:2014eia}, with the difference that here we replicate the theory (taking $n$ stacks of it), instead of putting the theory on a replicated manifold ($n$-sheeted Riemann surface).
We then get
\begin{equation}
	\Tr \slr{\rho_A(\tau)^n} = g_d^{1-n} \, \delta^{\Delta_{\cT_n}} \, \langle \cT_n(z) \rangle_\sM,
\end{equation}
where it is understood that the theory that lives on $\sM$ is now the $n$-replicated theory.

From this point, one can follow exactly the procedure done for one-point functions explained previously, obtaining that 
\begin{equation}
	\Tr \slr{\rho_A(t)^n} = 
	\begin{cases}
		g_a^{1-n} g_d^{1-n} (2 x / \epsilon')^{-\Delta_{\cT_n}} & t < x \\
		g_b^{1-n} g_d^{1-n} (2 x / \epsilon')^{-\Delta_{\cT_n}} & t > x .
	\end{cases}
\end{equation}
Finally, we recall that the R\'enyi entropies in the ground state of a CFT on a half-line, where one subsystem is adjacent to the boundary with boundary conditions (b.c.) \( a \) and \( d \) at the entangling surface, follow the scaling~\cite{Cardy:2016fqc}
\begin{equation}
	S^{(n)}_{A, \text{ground state}} =  \frac{c}{12} \frac{n+1}{n} \log \!\left( \frac{2x}{\epsilon'} \right) + \log (g_a) + \log (g_d).
\end{equation}
From this expression, we see that the R\'enyi entropies, like all one-point functions of scalar primaries, exhibit a light-cone behavior.
Keeping the subsystem fixed, the R\'enyi entropies transition from those of the initial ground state (with b.c. \( a \)) to those of the final ground state (with b.c. \( b \)), the transition occurring sharply when the subsystem \( A = [0,x] \) enters the light cone. The jump is given by
\begin{equation}\label{eq:entropy-jump}
	S^{(n)}_A(t) - S^{(n)}_A(0) = 
	\begin{cases}
		0 & \text{if } t < x \\
		\log \!\left( \frac{g_b}{g_a} \right) & \text{if } t > x.
	\end{cases}
\end{equation}
Since this relation holds for all R\'enyi entropies, it also holds for the entanglement entropy.
The result is independent of the \bc $d$ set on the entangling surface.

\section{Examples in the Ising model}
\label{sec:ising}
We specialize the general results of the previous section to the Ising CFT, which admits a simple lattice realization, which can be numerically simulated with Matrix Product States (MPS). 
In the Ising CFT there are three conformal boundary conditions: two {\it fixed} \bcs, $+$ and $-$, and one {\it free} \bc $f$ \cite{Cardy:1986gw}. We study the time evolution of the magnetization (or spin) operator $\sigma$, and of the entanglement entropy for the subsystem $A$, adjacent to the left boundary. We also study the time evolution of the $\bZ_2$ entanglement asymmetry~\cite{Vaccaro:2007drw, Ares:2022koq,Ferro:2023sbn} on $\rho_A$, which we will  introduce soon.

\paragraph{Quantum Ising model on an open chain}
We consider the critical quantum Ising model on an open chain of $L$ sites. 
The Hamiltonian with ``free'' \bcs on both endpoints is
\begin{equation}\label{eq:ising-ham-ff}
	H_{(f,f)} = - \sum_{j=1}^{L-1} X_j X_{j+1} - \sum_{j=1}^L Z_j.
\end{equation}
Fixed \bcs are set by adding an extra spin to the chain, which is polarized with infinite longitudinal field. For instance, the Hamiltonian for fixed \bcs on both endpoints is
\begin{equation}
	H_{(\pm, \pm)}^\mathrm{extra} = - \sum_{j=0}^{L} X_j X_{j+1} - \sum_{j=1}^L Z_j - h_L X_0 - h_R X_{L+1},
\end{equation}
and the \bc $+$ is obtained setting $h_L, h_R$ to $+\infty$, while $h_L, h_R = -\infty$ realizes the \bc $-$.
The Hilbert space of finite energy states is then $\ket{\pm} \otimes \lr{\bC^2}^{\otimes L} \otimes \ket \pm \cong (\bC^2)^{\otimes L}$. Since $X_0, X_{L+1} = \pm 1$, the Hamiltonian that acts on $ (\bC^2)^{\otimes L}$ is 
\begin{equation}\label{eq:ising-ham-fixed}
	H_{(\pm, \pm)} = - \sum_{j=1}^{L-1} X_j X_{j+1} - \sum_{j=1}^L Z_j \mp  X_1 \mp  X_L,
\end{equation}
thus effectively setting a boundary field with absolute value $1$ on a chain of length $L$.

Finally, we recall that the lattice magnetization operator $X_j$ flows to the CFT operator $\sigma$ (and its descendants, which are suppressed by higher powers of the lattice spacing).
We also recall that with the current normalization of the Hamiltonian in \cref{eq:ising-ham-ff,eq:ising-ham-fixed}, the velocity is $v = 2$. In order to match with the previous sections, where $v$ was assumed to be $1$, we will replace $t$ with $t/v$. 


\paragraph{Conformal data}
The bulk-boundary structure constants are~\cite{Runkel:1998he}
\begin{align}\label{eq:ising-bulk-bound-str-const}
	\cA_{\sigma, \bI}^+ &= - \cA_{\sigma, \bI}^- = \sqrt 2 ~, & \cA_{\sigma, \bI}^f &= 0,
\end{align}
and the values of the $g$-factors of the conformal \bcs are \cite{Affleck:1991tk}
\begin{align}\label{eq:g-factors-ising}
	g_+ = g_- &= 1 / \sqrt 2 ~,& g_f &= 1.
\end{align}

\paragraph{Periodicities and change of sign in the magnetization}
While the details of the simulations will be explained later, we anticipate one observation. Depending on the two boundary conditions, the time evolution of observables can be periodic with period $2L/v$ or $4L/v$. Moreover, when the period is $4L/v$, it is also observed that the magnetization profile changes sign after time $2L/v$. 
While we do not provide a complete analytic result for the time evolution of one point functions when the space manifold is a finite interval, it is possible to predict the occurrence (or absence) of a change of sign in the magnetization at time $t = 2L/v$ with a simple algebraic argument.

In the Ising CFT, there is a one to one correspondence between conformal boundary conditions and primary fields. We recall their relation, and the (chiral) conformal dimension of the primaries, which will be relevant later:
\begin{center}
	\begin{tabular}{ccc}
		b.c.& primary & $h$ \\
		\midrule
		$+$ & 1 & 0 \\
		$-$ & $\varepsilon$ & 1/2 \\
		$f$ & $\sigma$ & 1/16 \\
	\end{tabular}
\end{center}
The interval Hilbert space $\cH_{a,b}$ can be decomposed into irreducible modules of its Virasoro algebra, denoted by $\cV_c$. The multiplicities with which they appear are given by the fusion rules of the primaries under operator product expansion~\cite{Cardy:1989ir}
\begin{equation}
	\mathcal H_{a,b} = \bigoplus_{c} N_{a, b}^c \,  \cV_c.
\end{equation}
Denoting the Virasoro modules with the conformal dimension of the primary, we get
\begin{align}
	\cH_{+,+} &= \cH_{-,-} = \cV_0 & \cH_{f,f} &= \cV_0 \oplus \cV_{1/2} \notag \\
	\cH_{+,-} &= \cH_{-,+} =\cV_{1/2} & \cH_{f,+} = \cH_{f, -} &= \cV_{1/16}.
\end{align}

The Hamiltonian on the interval Hilbert space is 
\begin{equation}
	H_{a,b} = \frac{\pi}{L} \lr{ L_0 - \frac{c}{24} },
\end{equation}
and the time evolution operator is $U(t) = \exp \lr{-i t H} $. The spectral structure of $L_0$ determines certain periodicities in the time evolution: the eigenvalues of $L_0$ on the irreducible module $\cV_c$ take values in $h_c + \bN$, so that at time $t=2L$ all the states in a given module evolve by the same phase. Thus,
\begin{equation}
	U(2L) = e^{i \pi c /12}  \times \, \bigoplus_a e^{i 2\pi h_a} \, \id_{\cV_a} \, ,
\end{equation}
where $\id_{\cV_a}$ denotes the identity map on the module ${\cV_a}$. 
At time $t=2L$, with \bcs $(f,f)$, one obtains
\begin{equation}
	U(2L) = e^{i \pi c /12} 
	\begin{pmatrix}
		\id_{\cV_0}  & 0 \\
		0 & - \id_{\cV_{1/2}}
	\end{pmatrix}.
\end{equation}
Now recall that the states in $\cV_0$ are $\bZ_2$-even, while the states in $\cV_{1/2}$ are $\bZ_2$-odd. Then, up to a global phase, $U(2L)$ is equal to the $\bZ_2$ spin-flip operator, hence
\begin{equation}
	\label{eq:magnetization-change-sign}
	U(2L)  \, \hat \sigma(x) \, U(2L)^\dagger = - \hat \sigma(x)
\end{equation}
i.e. the magnetization profile changes sign at time $t=2L$. 
At time $t = 4L$, one has $U(4L) = e^{i\pi c / 6} \mathds{1}_{\cH_{f,f}}$, so the time evolution is periodic with period $4L$.

\paragraph{}
When any other choice of \bcs is taken, the Hilbert space is made up of a single Virasoro module. Then the magnetization does not change sign, as $U(2L)$ becomes proportional to the identity and $U(2L) \, \hat \sigma(x) U(2L)^\dagger = \hat \sigma(x)$.

\subsection{Numerical simulations}
We perform matrix product states (MPSs) simulations to confirm the analytical results of \cref{sec:time-evol-1pt-funcs}. Simulations are performed with the Tenpy library~\cite{tenpy} on a chain of $150$ sites. Lattice spacing is assumed to be the unit length, so $L=150$. 
The ground state is obtained with the DMRG algorithm, with a maximum bond dimension of $30$, and then evolved in time with the TEBD algorithm, with a time step $dt=0.05$.
At every time, we study the expectation value of the local magnetization operator $X_j$, and of the entanglement entropy on the interval $[0, x]$.
We also study the second R\'enyi asymmetry for the $\bZ_2$ group, generated by $\prod_{j=1}^L Z_j$ describing the spin-flip action.

\paragraph{Entanglement asymmetry}
We study numerically, on a lattice realization, the second R\'enyi asymmetry \cite{Ares:2022koq}, which measures how much a state breaks a given symmetry. In the following, we recall its definition.  Let $G$ be a finite group, represented on the Hilbert space with unitary operators $U_g, \, g \in G$. We assume the group representation to be on-site and thus to factorize over a spatial bipartition $A \cup B$ as $U_g = U_{A,g} \otimes U_{B,g}$, where $U_A$ is an operator on $\cH_A$ and $U_B$ on $\cH_B$.
The $n$-th R\'enyi asymmetry is defined as
\begin{equation}
	\Delta S^{(n)}_A = \frac{1}{1-n} \log \frac{\Tr \slr{ \lr{\rho_{A,\sS}}^n}} {\Tr \slr{ \lr{\rho_{A}}^n}},
\end{equation}
where $\rho_{A,\sS}$ is the \emph{symmetrized} reduced density matrix
\begin{equation}
	\rho_{A, \sS} = \frac{1}{|G|} \sum_{g \in G} U_{A,g} \rho_A U_{A,g}^\dagger.
\end{equation}
The \emph{entanglement asymmetry} can be obtained by taking the analytic continuation of the R\'enyi asymmetry in the variable $n$ and taking the limit $n\to 1$. In the following, we will study the second R\'enyi asymmetry as a more accessible proxy for the entanglement asymmetry. 
In our case, $G = \bZ_2 = \{1, \eta\}$, and $\rho_{A,\sS} = \frac{1}{2} (\rho_A + \eta_A \rho_a \eta_A)$, with $\eta_A = \prod_{j \in A} X_j$.

\begin{figure}
	\centering
	\includegraphics[width=0.8\textwidth]{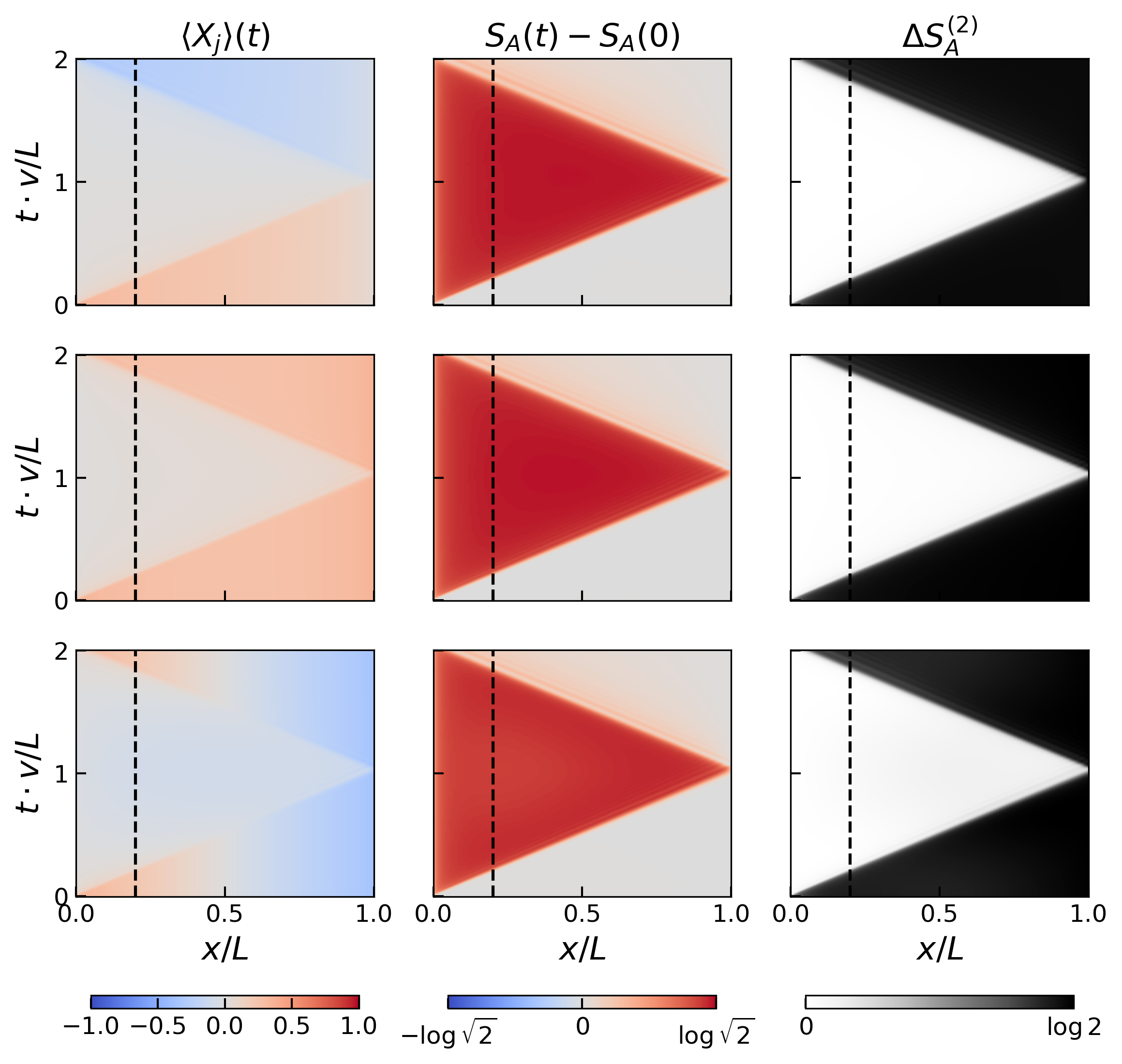} 
	
	\includegraphics[width=0.75\textwidth]{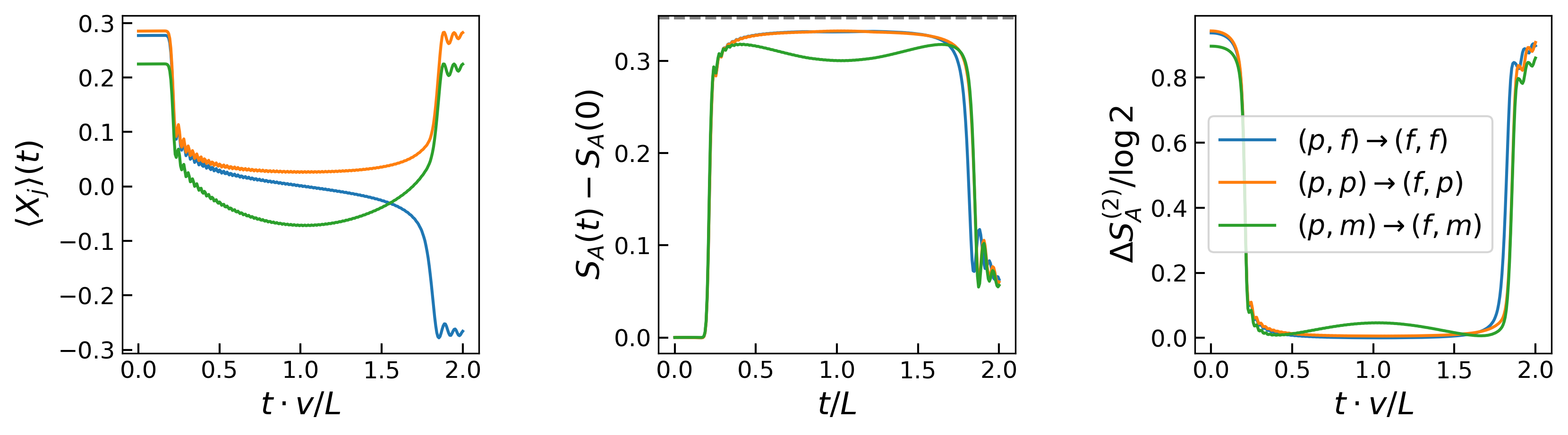}
	\caption{
	\textbf{Top}: Space-time plot of the quench $(+, b) \to (f, b)$. Every row is a different quench, where $b$ is fixed respectively to $f, +, -$. 
	Columns are respectively: magnetization, jump in the entanglement entropy, second R\'enyi asymmetry on $A=[0,x]$.
	The predicted light-cone behavior is observed in the magnetization and in the entanglement entropy.
	\textbf{Bottom}: The same observables, on a fixed site at $x/L = 1/5$ (dashed line in top figures), as a function of time. 
	}
	\label{fig:p2f}
\end{figure}

\paragraph{Quench $\boldsymbol{+ \to f}$}
The quench $+ \to f$ is performed by preparing the ground state  with \bc $+$ on the left edge, then turning off the boundary field and evolving in time with the free boundary condition.
We choose a boundary condition $b \in \{+,-,f\}$ on the right boundary and leave it unchanged during the quench. We repeat the simulation for the three possible values of $b$. 
For times and positions much smaller than $L$, the system behaves as if it were semi-infinite, so the results of \cref{sec:time-evol-1pt-funcs} apply. From \eqref{eq:ising-bulk-bound-str-const}, we expect the magnetization to vanish inside the light-cone, and from \eqref{eq:entropy-jump} we expect the entropy to have a positive jump of $\log(\sqrt 2)$. These are observed and reported in \cref{fig:p2f}.
The change in sign of the magnetization at time $t=2L$ predicted by \eqref{eq:magnetization-change-sign} is observed when the time evolution is with \bcs $(f,f)$, while it is absent for other choices of \bcs.

The $\bZ_2$-asymmetry on $A=[0,x]$ also displays a lightcone behavior. At $t=0$ it starts from a positive value, approximately $\log 2$, with power-law corrections, as obtained in \cite{Fossati:2024ekt,Ferro:2023sbn}.
It then becomes zero when $[0,x]$ fully enters the lightcone, at $t=x/v$, thus showing that the symmetry is locally restored. For $vt\in [L, 2L]$, it goes back to its initial value. 
The asymmetry is always periodic over a period of $2L/v$: for \bcs $(f,f)$, it holds that $\rho(t + 2L/v) = \eta \rho(t) \eta$, which implies $\Delta S^{(n)}_A(\rho_A(t + 2L/v)) = \Delta S^{(n)}_A(\rho_A(t))$.

\begin{figure}
	\centering
	\includegraphics[width=0.8\textwidth]{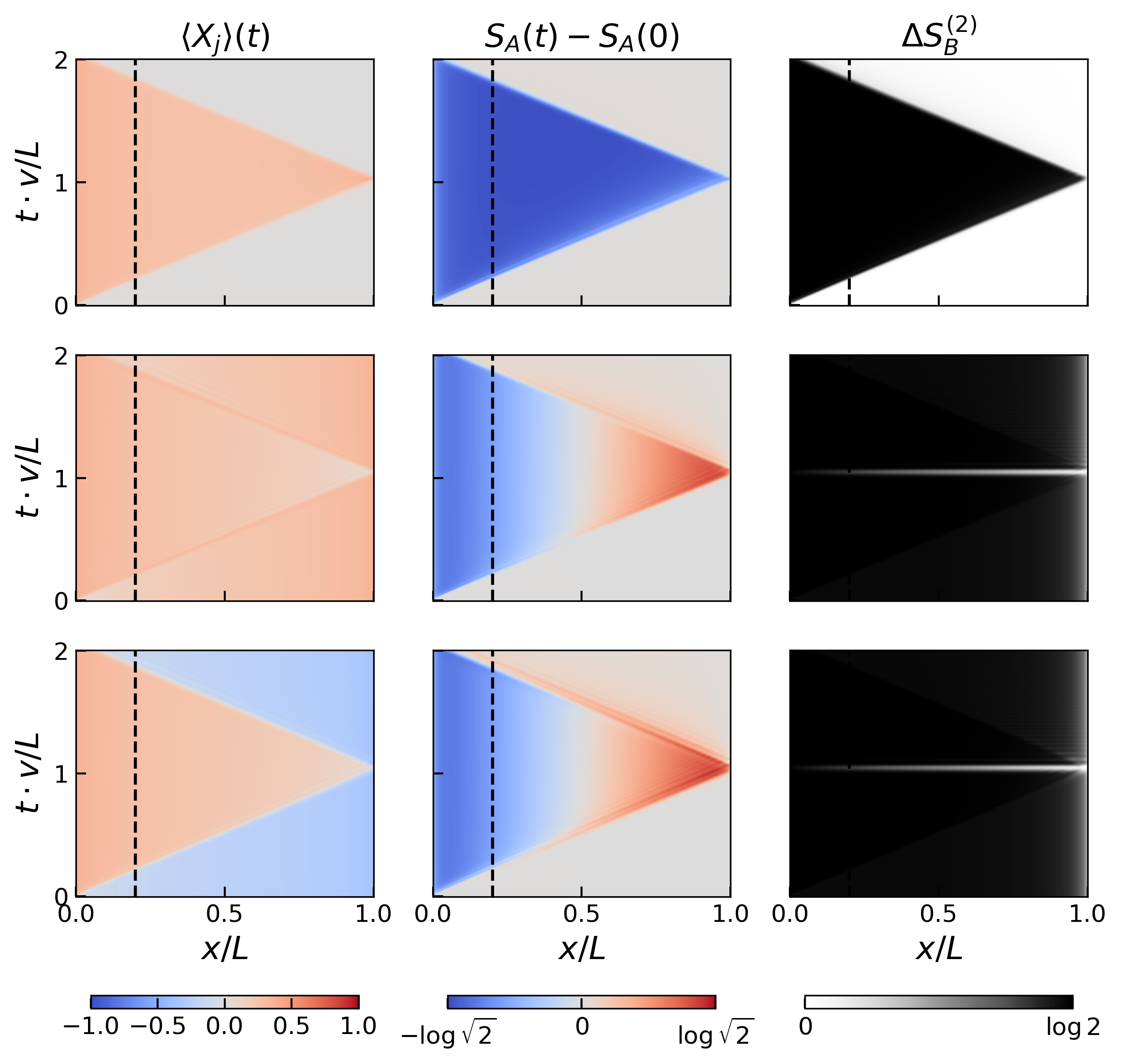}
	
	\includegraphics[width=0.75\textwidth]{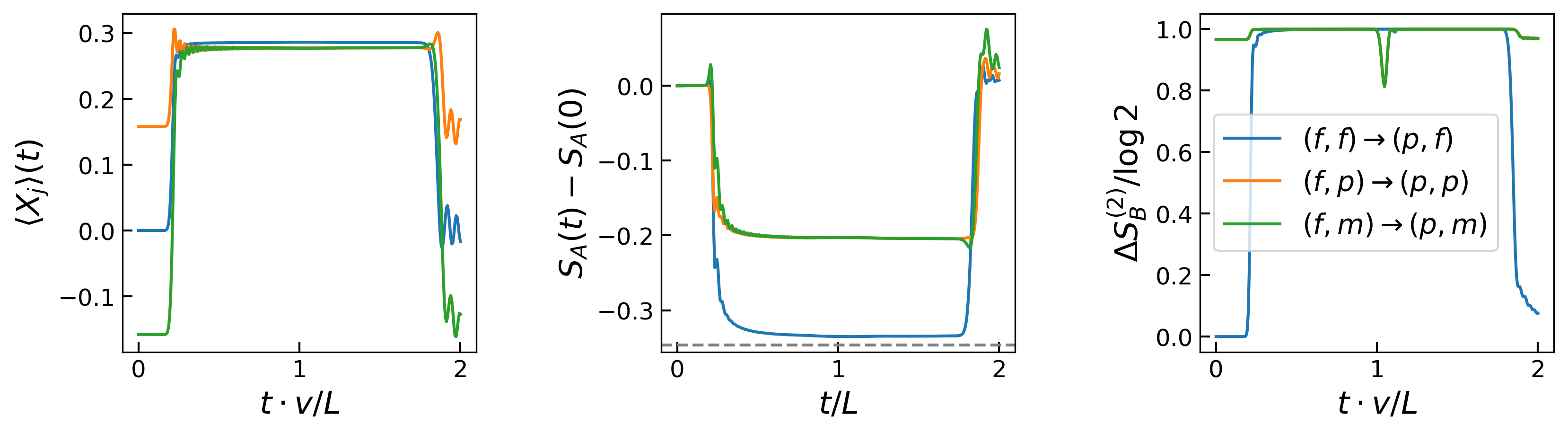}
	\caption{\textbf{Top}: Space-time plots of the quench $(f, b) \to (+, b)$. Every row is a different quench, where $b$ is fixed respectively to $f, +, -$. 
		Columns are respectively: magnetization, jump in the entanglement entropy, second R\'enyi asymmetry on the right subsystem $B=[x,L]$.
		\textbf{Bottom}: The same observables, on a fixed site at $x/L = 1/5$ (dashed line in top figures), as a function of time. }
	\label{fig:f2p}
\end{figure}

\paragraph{Quench $f \to +$}
We then study the opposite quench, in which the left boundary condition is changed from free to $+$. We keep the right boundary condition $b$ fixed, for the three possible values of $b$. We report the results in \cref{fig:f2p}.

As expected, for small values of $x$, the magnetization becomes positive in the light-cone. Its value at initial time would be zero on an infinite chain, but is affected by the boundary condition $b$ on the right at finite size. 
The entropy is expected to have a negative jump of $-\log \sqrt 2$, which is clearly observed when the right \bc is $f$. When the right \bc is $+$ or $-$ the jump is not uniform: it is negative for $x/L \sim 0$ and positive for $x/L \sim 1$. The jumps are smaller than $\log \sqrt 2$ in absolute value, and we believe that they converge to such values for large enough $L$. 

The $\bZ_2$ asymmetry is  reported for the right subsystem $B=[x,L]$, because the one on $A$ is almost always equal to $\log 2$, and does not display interesting dynamics. For the right boundary condition $f$, the entanglement asymmetry vanishes at $t=0$, reflecting the fact that the initial state is symmetric. Then, as time passes, the symmetry remains broken in $B$ if $x > t/v$, hence the light-cone behavior. When the right boundary condition is fixed, the asymmetry is non-zero even at initial time. 

At time $t=L/v$, the asymmetry is approximately linear in $x$. This behavior is typical of the ground state $(+,f)$ (or $(-,f)$)~\cite{Fossati:2024ekt}, and thus suggests that the time-evolved ground state with boundary condition $(f,+)$ at time $L/v$ \emph{is} the ground state with \bcs $(f,+)$. This possibility is also compatible with the spatial profile of the magnetization and of the entanglement entropy at that time.

\begin{figure}
	\centering
	\includegraphics[width=0.5\textwidth]{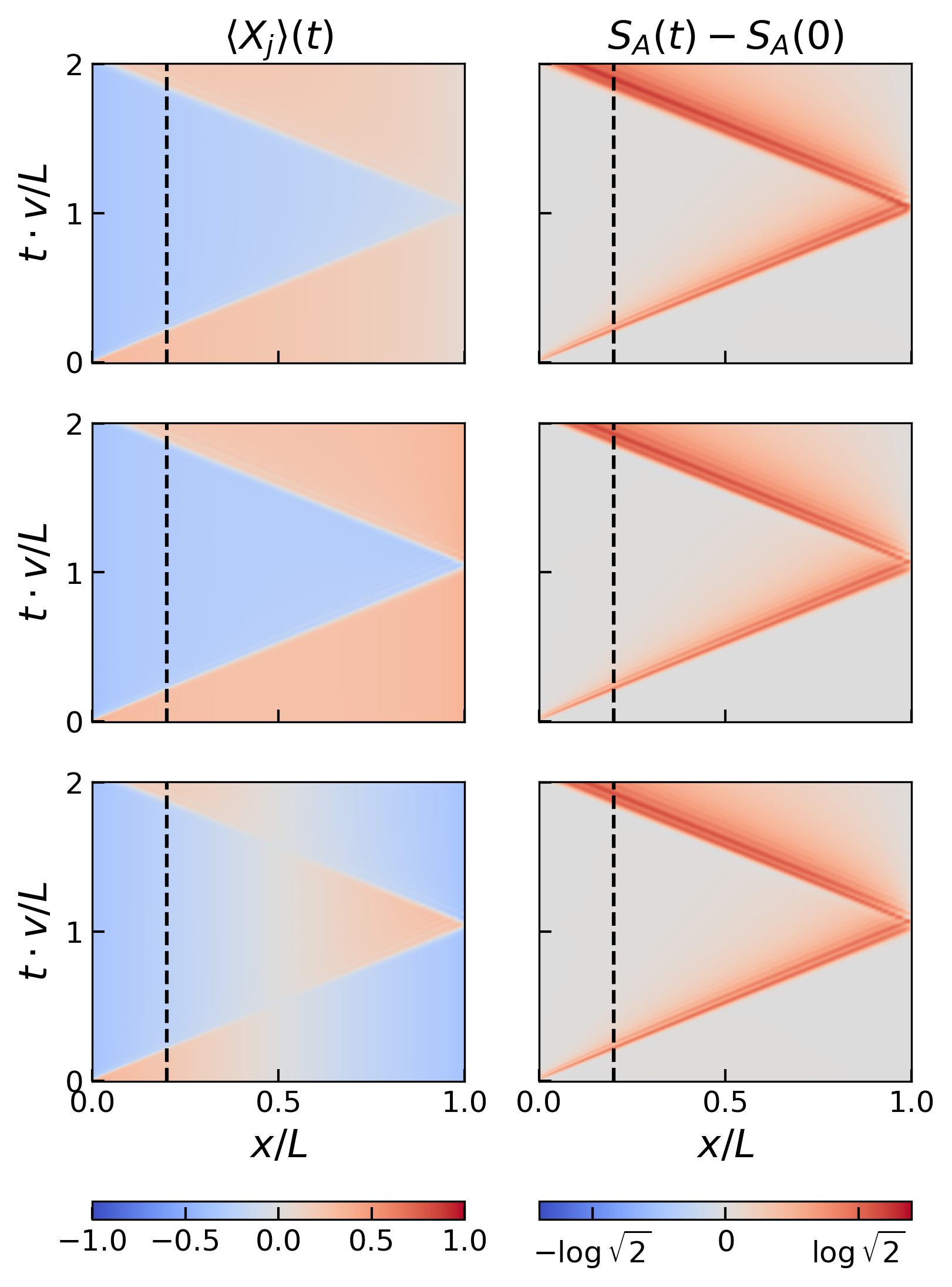}
	
	\includegraphics[width=0.4\textwidth]{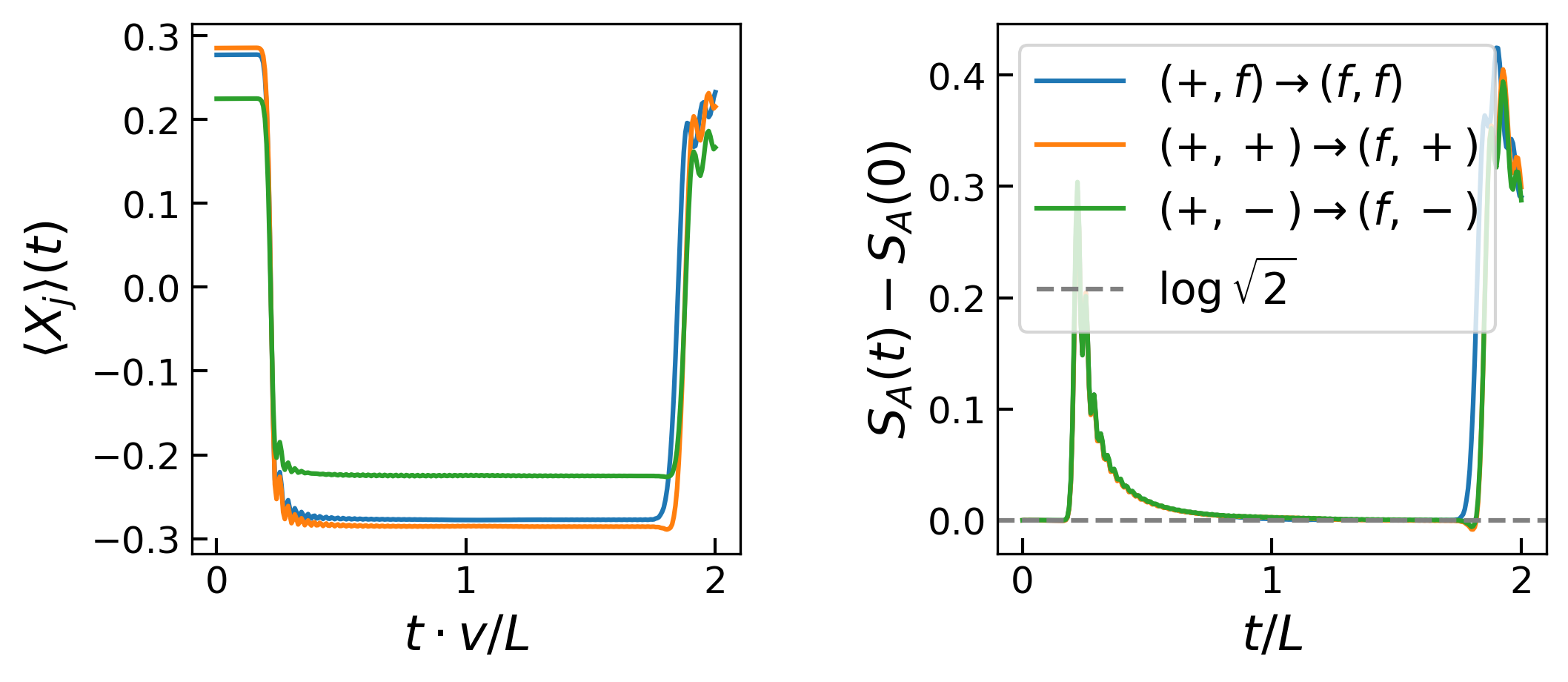}
	\caption{\textbf{Top}: Spacetime plot for the $(+, b) \to (-, b)$ quench. Every row is a different quench, where $b$ is fixed respectively to $f, +, -$. 
	Columns display magnetization and variation in the entanglement entropy.
\textbf{Bottom}: The same observables, on a fixed site at $x/L = 1/5$ (dashed line in top figures), as a function of time .	}
	\label{fig:p2m}
\end{figure}

\paragraph{Quench $+ \to -$}
Lastly, we consider the boundary quench in which the boundary field is initially non-zero and undergoes a change of sign at $t=0$.
Here, the magnetization close to the left edge is expected to change sign, while the entropy should not display any jump, because $g_a = g_b$. The results are shown in \cref{fig:p2m}. Magnetization does indeed change sign when crossing the lightcone. The entanglement entropy has a spike in the correspondence of the lightcone, restoring to its initial value right after that, as expected. 
The second R\'enyi asymmetry is almost uniformly equal to $\log 2$, because the \bcs break the symmetry both before and after the quench. Hence, we do not display it here.

\section{Conclusion and outlook}

In this work, we have studied boundary quenches in $(1+1)$-dimensional conformal field theories. We have shown that, on the half-infinite line, when the boundary condition is abruptly changed from $a$ to $b$, one-point functions coincide with those of the ground state with boundary condition $a$ in the space-like region of the quench event, and with those of the ground state with boundary condition $b$ in the time-like region (see Eq.~\eqref{eq:one-pt-func-result}). As a consequence, the entanglement entropy of a bipartition with one subsystem adjacent to the boundary exhibits a jump of $\log(g_b/g_a)$ when the entire subsystem becomes time-like separated from the quench event (see Eq.~\eqref{eq:entropy-jump}).

In addition to ultracold atoms and trapped ions, our CFT results are also expected to have direct implications to solid state systems with Kondo correlations~\cite{PhysRevB.48.7297} which can be switched on and off suddenly~\cite{Ma:2025tol}.

Several directions deserve further investigation. A natural extension is to study analytically boundary quenches on a finite interval, where finite-size effects and multiple reflections from the boundaries are expected to produce richer dynamics. In particular, it would be interesting to understand whether, at special times such as $t=L/v$, the time-evolved state can again be identified with the ground state with other conformal boundary conditions. Another important direction is the study of two-point and higher-point correlation functions, which should provide a more complete characterization of the post-quench dynamics.

Finally, it would be interesting to understand whether the quasiparticle picture for global quenches~\cite{Calabrese:2005in} can be extended to accommodate the boundary quenches considered here. In its standard formulation, the quasiparticle picture assumes that the entanglement is transported by pairs of entangled quasiparticles emitted throughout the system at the quench. However, it is not immediately clear how this framework could be adapted to yield quantitative predictions for boundary quenches (or, more generally, for local quenches).
Developing such a quasiparticle interpretation would be highly desirable, as it could naturally be generalized to interacting integrable models, where it has proved remarkably successful~\cite{Alba:2017ekd,calabrese_ln}, and would therefore extend the present results well beyond the critical regime.

\section*{Acknowledgements}
ES acknowledges support from the European Research Council (ERC) under the European Union Horizon 2020 research and innovation programme under grant agreement No. 951541. PC thanks the European Commission through the ERC-AdG grant MOSE No.\ 101199196.
	
\bibliographystyle{ytphys}
\bibliography{bibliography}
\end{document}

%% file: setup.tex
\usepackage{amsmath,amssymb,amsfonts,amsthm,mathrsfs,mathtools,cases}
\usepackage{dsfont}
\usepackage{authblk}
\usepackage[colorlinks=true,linkcolor=blue, citecolor=blue, bookmarks]{hyperref}
\usepackage{cleveref}
\usepackage{microtype}
\usepackage{comment}
\usepackage{graphicx,adjustbox} 
\usepackage{subfig}
\usepackage{wrapfig}
\usepackage[font=small,labelfont=bf, margin=0pt, width=\linewidth]{caption}
\usepackage[usenames,dvipsnames]{xcolor}  
\usepackage{booktabs}
\usepackage{braket}
\usepackage[nottoc]{tocbibind}
\usepackage{nicematrix}
\usepackage{tikz}
\usepackage{pgfplots}
\usepackage{cite}
\pgfplotsset{compat=1.18}
\usepackage{enumitem}

\usetikzlibrary{matrix}
\usetikzlibrary{decorations.markings}
\usetikzlibrary{patterns}
\usetikzlibrary{arrows.meta}
\pgfmathsetmacro\mathaxis{-height("$\vcenter{}$")}

\numberwithin{equation}{section}

\newcommand{\bc}{boundary condition }
\newcommand{\bcs}{boundary conditions }

\newcommand{\cA}{\mathcal{A}}

\newcommand{\cF}{\mathcal{F}}

\newcommand{\cH}{\mathcal{H}}

\newcommand{\cO}{\mathcal{O}}

\newcommand{\cT}{\mathcal{T}}

\newcommand{\cV}{\mathcal{V}}

\newcommand{\bC}{\mathbb{C}}

\newcommand{\bI}{\mathbb{I}}

\newcommand{\bN}{\mathbb{N}}

\newcommand{\bR}{\mathbb{R}}

\newcommand{\bZ}{\mathbb{Z}}

\newcommand{\sM}{\mathsf{M}}
\newcommand{\sH}{\mathsf{H}}

\newcommand{\sS}{\mathsf{S}}

\newcommand{\id}{\mathrm{id}}

\DeclareMathOperator{\End}{End}

\DeclareMathOperator{\sign}{sgn}

\DeclareMathOperator{\Tr}{Tr}

\newcommand{\lr}[1]{\left( {#1} \right)}
\newcommand{\slr}[1]{\left[{#1} \right]}

\newcommand{\gs}{\mathrm{gs}}

\renewcommand{\Re}{\operatorname{Re}}
\renewcommand{\Im}{\operatorname{Im}}
\newcommand{\abs}[1]{\left | #1 \right|}

\crefname{figure}{Fig.}{Figs.}
\crefname{equation}{Eq.}{Eqs.}
\crefname{section}{Sec.}{Secs.}
\crefname{appendix}{Appendix}{Appendices}
  
\colorlet{darkerblue}{MidnightBlue!70!black}
\colorlet{lightblue}{blue!70!white}

\hypersetup{
    colorlinks,
    linkcolor={darkerblue},
    citecolor={darkerblue},
    urlcolor={darkerblue},
    backref=true
}

\newcommand{\tikzmath}[2][]{\vcenter{\hbox{\begin{tikzpicture}[#1]#2
	\end{tikzpicture}}}
}

\usepackage[framemethod=TikZ]{mdframed}
